\documentclass[]{aastex631}

\accepted{Oct 13, 2021}

\submitjournal{APJ}

\shortauthors{Zhao et al.}
\graphicspath{{./}{figures/}}

\begin{document}

\title{Analysis of Magnetohydrodynamic Perturbations in Radial-field Solar Wind from Parker Solar Probe Observations}


\author[0000-0003-4268-7763]{S. Q. Zhao}
\affiliation{Deutsches Elektronen Synchrotron (DESY) Platanenallee 6, D-15738, Zeuthen, Germany}
\affiliation{ Institut für Physik und Astronomie, Universität Potsdam, D-14476, Potsdam, Germany}

\author[0000-0003-2560-8066]{Huirong Yan}
\affiliation{Deutsches Elektronen Synchrotron (DESY) Platanenallee 6, D-15738, Zeuthen, Germany}
\affiliation{ Institut für Physik und Astronomie, Universität Potsdam, D-14476, Potsdam, Germany}

\author[0000-0003-1778-4289]{Terry Z. Liu}
\affiliation{Department of Earth, Planetary, and Space Sciences, University of California, LA, USA}

\author[0000-0003-2981-0544]{Mingzhe Liu}
\affiliation{LESIA, Observatoire de Paris, Université PSL, CNRS, Sorbonne Université, Université de Paris, place Jules Janssen, 92195, Meudon, France}

\author[0000-0002-9201-5896]{Mijie Shi}
\affiliation{Centre for mathematical Plasma Astrophysics, Department of Mathematics, KU Leuven, B-3001 Leuven, Belgium}

\correspondingauthor{Huirong Yan}
\email{huirong.yan@desy.de}
\begin{abstract}
We report analysis of sub-Alfvénic magnetohydrodynamic (MHD) perturbations in the low-$\beta$ radial-field solar wind employing the Parker Solar Probe spacecraft data from 31 October to 12 November 2018. We calculate wave vectors using the singular value decomposition method and separate MHD perturbations into three eigenmodes (Alfvén, fast, and slow modes) to explore the properties of sub-Alfvénic perturbations and the role of compressible perturbations in solar wind heating. The MHD perturbations show a high degree of Alfvénicity in the radial-field solar wind, with the energy fraction of Alfvén modes dominating ($\sim$45\%-83\%) over those of fast modes ($\sim$16\%-43\%) and slow modes ($\sim$1\%-19\%). We present a detailed analysis of a representative event on 10 November 2018. Observations show that fast modes dominate magnetic compressibility, whereas slow modes dominate density compressibility. The energy damping rate of compressible modes is comparable to the heating rate, suggesting the collisionless damping of compressible modes could be significant for solar wind heating. These results are valuable for further studies of the imbalanced turbulence near the Sun and possible heating effects of compressible modes at MHD scales in low-$\beta$ plasma.

\end{abstract}


\section{Introduction} \label{sec:intro}
Plasma turbulence appears ubiquitous and plays a crucial role in various astrophysical processes, such as the solar wind heating and acceleration \citep{Bandyopadhyay2020}, scattering of cosmic rays \citep{Yan2015}, turbulent heating in galaxy clusters \citep{Zhuravleva2014}, and star formation \citep{Federrath2018}. Turbulence is typically characterized by a broadband spectrum of perturbations, energy transmission spanning a vast range of temporal and spatial scales, highly chaotic but self-similar motions within the inertial range. The solar wind, a plasma flow originating from the Sun and continuously blowing into the interplanetary space, provides an excellent laboratory for studying plasma turbulence at magnetohydrodynamic (MHD) and sub-ion-kinetic scales \citep{Dobrowolny1980, Verscharen2019}. MHD perturbations can be decomposed into three eigenmodes: Alfvén, fast, and slow modes \citep{Glassmeier1995, Cho2003}. Using the term 'mode' in this study, we refer to the carriers of turbulent perturbations in wave turbulence rather than classical linear waves \citep{Cho2003, Verscharen2019}. The mode composition affects almost all turbulence dynamics and the mechanisms of solar wind heating \citep{Suzuki2006, Cranmer2012, Makwana2020}. Clarifying the mode composition and the properties of each mode can help us further understand astrophysical mysteries, e.g., corona heating; transport of particles \citep{Chandran2005, Yan2008}.

The MHD mode composition has been extensively investigated through simulations and satellite observations \citep{Cho2003, Makwana2020, Chaston2020, Zhu2020}. Simulations of MHD turbulence found that different modes cascade differently. The cascade of Alfvén and slow modes is anisotropic, mainly in the direction perpendicular to the local background magnetic field, whereas fast modes tend to show isotropic cascade \citep{Cho2003, Makwana2020}. Furthermore, satellite observations with the mode composition diagnosis \citep{Glassmeier1995} show that anti-sunward propagating Alfvén modes dominate mode composition, and energy fraction enhancements of compressible modes are associated with the background magnetic field variations \citep{Chaston2020, Zhu2020}.

The compressible modes affect the compressibility of solar wind turbulence and thus influence other turbulence properties \citep{Gary1986, Chen2020}. Magnetic compressibility, defined as $C_{BB}=(\delta|\mathbf{B}|/|\delta\mathbf{B}|)^2$, is observed to increase with the heliocentric distance \citep{Bavassano1982, Chen2020, Andres2021}. \citet{Howes2012} indicated that compressible components of solar wind turbulence at the inertial range mainly result from the kinetic slow mode. Additionally, \citet{Chen2020} suggested that magnetic compressibility depends on plasma $\beta$ and slow-mode perturbations with the assumption of $\delta{B}_{\perp}$ from Alfvén modes and $\delta|\mathbf{B}|\approx\delta{B}_{\parallel}$ from slow modes. Then a straightforward but critical question is whether the assumption of the magnetic compressible component ($\delta{B}_{\parallel}$) from slow modes is still valid in quiet radial-field solar wind turbulence \citep{Bale2019}. The normalized radial magnetic field is required to satisfy $|B_R|/|\mathbf{B}|>0.8$ in this study, where $B_R$ represents the magnetic field along the direction of the Sun to the spacecraft in radial tangential normal (RTN) coordinates. To investigate the possible origins of magnetic and density compressibility, we separate the MHD perturbations into three eigenmodes using the mode decomposition method \citep{Cho2003} to explore the role of each mode in kinetic, magnetic, and density power spectra.

The collisionless damping of plasma waves plays a critical role in plasma heating \citep{Suzuki2006, Zhuravleva2014}. In the MHD regime, collisionless damping of compressible modes is widely considered a possible heating mechanism \citep{Porter1994,  Spanier2005, Petrosian2006, Kumar2006}. Therefore, another critical question, which is not well understood, is about the possible heating effects of each mode. Alfvén modes are non-compressive and can only be damped ohmically or by shear viscosity. Thus, Alfvén modes are weakly damped in a plasma with Maxwellian distributions, making limited contributions to plasma heating in the linear regime. By contrast, compressible magnetosonic modes (fast and slow modes) are prone to intense collisionless damping owing to wave-particle interactions, providing an efficient energy conversion between waves and plasmas \citep{Ginzburg1961, Barnes1967}. The compressible MHD perturbations are typically interpreted as a mixture of fast magnetosonic waves and pressure-balanced structures (PBS). The PBS is characterized by an anti-correlation of thermal and magnetic pressure, similar to slow modes \citep{Bruno2013, Howes2012}. Therefore, mode decomposition provides a valuable tool for quantitatively analyzing the possible role of each mode in plasma heating.

The Parker Solar Probe (PSP) mission is well situated to investigate the properties of the near-Sun turbulence with the nearest heliocentric distance of ~0.17 AU during its first encounter. This study applies the mode decomposition method \citep{Cho2003} to in-situ spacecraft observations to study sub-Alfvénic MHD perturbations in the solar wind for the first time. This new mode decomposition method makes it possible to quantitively analyze the role of each MHD mode in kinetic, magnetic, and density power spectra and the contributions of each mode on magnetic and density compressibility. Moreover, we determine the collisionless damping of each mode to reveal the role of compressible modes in solar wind heating. The outline of this paper is as follows. Section \ref{sec:data} presents data sets, search criteria, and analysis methods used in this study. Section \ref{sec:observations} offers a representative case to analyze the mode composition of the MHD perturbations and the possible role compressible modes play in solar wind heating. In Section 
\ref{sec:Discussion} and \ref{sec:Summary}, we discuss and conclude our results.

\section{Data and Mode Decomposition}\label{sec:data}

\subsection{Data} \label{subsec:data_base}
We utilize data measured by NASA’s PSP mission \citep{Fox2016} during its first perihelion. We analyze magnetic field data from the fluxgate magnetometer (MAG; \citet{Bale2016}), proton parameters with a ~0.874 s resolution from the Solar Probe Cup (SPC; \citet{Kasper2016}), and electron parameters with a 7 s resolution deduced from the simplified quasi-thermal noise (QTN) method with observations from RFS/FIELDS \citep{Moncuquet2020}. The electric field data at MHD scales are calculated by $\mathbf{E}=-\mathbf{V_p}\times{\mathbf{B}}$, where $\mathbf{V_p}$ is the proton bulk velocity and $\mathbf{B}$ is the magnetic field. The observed parameters consist of the ensemble average background field and a fluctuating field, i.e., $\mathbf{B}=\mathbf{B_0}+\mathbf{b}$, $\mathbf{V_p}=\mathbf{V_0}+\mathbf{v}$, $\mathbf{E}=\mathbf{E_0}+\mathbf{e}$, $\mathbf{N_p}=\mathbf{N_0}+\mathbf{n}$. The  magnetic field perturbation $\mathbf{b}$, presented in Alfvén speed units, is normalized by $\sqrt{\mu_0\rho_0}=\sqrt{\mu_0m_pN_0}$, where $\mu_0$ is the vacuum permeability, $\rho_0$ is the mean proton mass density, and $m_p$ is the proton mass. The ensemble average in this study is represented by a time average over each 250 s interval.

\subsection{Data selection criteria} \label{subsec:criteria}
The first PSP encounter spans from 31 October to 12 November 2018, covering the heliocentric distance between 0.166 and 0.277 AU. We search for events that satisfy three criteria: (1) Sub-Alfvénic ($\mathbf{v}\ll\mathbf{V_A}$ and $\mathbf{b}\ll\mathbf{B_0}$ ) perturbations, where $\mathbf{V_A}$ is the Alfvén speed. Under such a condition, the MHD perturbations can be considered as a pure superposition of linear modes. Under such a condition, the nonlinear terms $(v^2,b^2)$ are much less than the linear terms $(\mathbf{V_A\cdot v},\mathbf{B_0\cdot b})$, and thus perturbations can roughly be considered as a pure superposition of three MHD modes. (2) Large normalized radial magnetic field ($|B_R|/|\mathbf{B}|>0.8$). In the radial-field solar wind, the small (quasi-parallel/quasi-antiparallel) flow-to-field angles $\theta_{BV}$ indicate wave vectors along the field larger than that across it $k_\parallel \gg k_\perp\sim0 $. Thus, these low-amplitude and field-aligned perturbations exhibit more wave-like characteristics than turbulence \citep{Belcher1971, Bale2019}. (3) High electromagnetic planarity $>$ +0.8 \citep{Santolik2003} means the validity of planarity assumption, guaranteeing that small perturbations can be written as Fourier components $(e^{i(\mathbf{k\cdot x}-\omega t)})$ in the linearization of MHD equations. It is the foundation both of the singular value decomposition (SVD) method and the mode decomposition method. (4) The event duration should be longer than 20 minutes to ensure the measurements of low-frequency signals. During the first PSP encounter, a total of 15 events are identified and listed in Table \ref{tab:messier}. The representative case analyzed in Section 3 corresponds to event \#15 in the table and the shaded area in Figure \ref{fig:Figure1}.

\subsection{The measurements of the dispersion relations} \label{subsec:SVD}
To measure the observed dispersion relations, we first determine wave vectors $\mathbf{k}$ using the SVD method of \citet{Santolik2003}. The SVD method holds under the assumption of a single plane wave. This technique provides a mathematical approach to obtain the frequency-time spectrograms of $\mathbf{k}$ through solving the linearized Faraday's law:  $\mathbf{k}\times\mathbf{E}(\omega_{sc},t)=\omega_{sc}\mathbf{B}(\omega_{sc},t)$, where complex matrices of electric and magnetic field ($\mathbf{E}(\omega_{sc},t)$ and $\mathbf{B}(\omega_{sc},t)$) are calculated through the Morlet-wavelet transform \citep{Grinsted2004} and $\omega_{sc}=2\pi f_{sc}$ is the observed frequency in the spacecraft frame. Then, according to the Doppler shift, the observed frequency in the plasma flow frame can be obtained by $\omega_{pf}=\omega_{sc}-<\mathbf{k}>\cdot <\mathbf{V}_p>$, where $< >$ represents the time average. The observed dispersion relations will be compared with the theoretical ones obtained from the mode decomposition method, as detailed below.

\subsection{Mode decomposition methods} \label{subsec:decompose}
We utilize \citet{Cho2003} method to decompose the MHD perturbations into three eigenmodes: Alfvén, slow, and fast modes. The three modes share the same wave vector. We first use $\mathbf{k}$ and $\mathbf{B_0}$ to build a new coordinate: $\mathbf{\xi}=\xi_{\perp}\hat{\mathbf{k}}_{\perp}+\xi_{\parallel}\hat{\mathbf{k}}_{\parallel}+\xi_{\phi}\hat{\mathbf{\phi}}$ in wavevector space, where displacement vectors $\mathbf{\xi}$ are defined through $\frac{\partial \mathbf{\xi}}{\partial t}=\mathbf{v_k}$, and $\mathbf{v_k}$ is the phase-space fluctuating velocity.
$\hat{\mathbf{k}}_{\perp}(\perp\mathbf{B_0})$, $\hat{\mathbf{k}}_{\parallel}(\parallel\mathbf{B_0})$, $\hat{\mathbf{\phi}}=\hat{\mathbf{k}}_{\perp}\times\hat{\mathbf{k}}_{\parallel}$ are unit vectors along the orientations of coordinate axes. The velocity perturbations do not comply with one single mode, i.e., shear Alfvén mode.

The time series of velocity perturbations is transformed into frequency space by fast Fourier transform (FFT). We transform velocity perturbation ($\mathbf{v_f}$) in the frequency domain into ($\mathbf{v_k}$) using the relationship between $k$ and $f$ determined by the SVD method, which connects temporal and spatial space. Because the velocity perturbation of each mode is along respective displacement vectors at each wavevector scale, ($\mathbf{v_k}$) is projected onto the corresponding $\mathbf{\xi}$ (i.e., $\mathbf{\xi_{Alfven}}$, $\mathbf{\xi_{slow}}$, $\mathbf{\xi_{fast}}$, see Figure \ref{fig:Figure2}(a)). The velocity perturbation, magnetic field, and density of each mode are given by
\begin{equation}\label{1}
\mathbf{v_k}=\mathbf{v_{k,Alfven}}+\mathbf{v_{k,slow}}+\mathbf{v_{k,fast}},
\end{equation}

\begin{equation}\label{2}
b_{k,Alfven}=B_0 \frac{v_{k,Alfven}}{V_A}; 
b_{k,slow}=B_0 \frac{v_{k,slow}}{c_{slow}}|\mathbf{\hat{B}_0\times\hat{\xi}_{slow}}|; 
b_{k,fast}=B_0 \frac{v_{k,fast}}{c_{fast}}|\mathbf{\hat{B}_0\times\hat{\xi}_{fast}}|;
\end{equation}

\begin{equation}\label{3} 
N_k=N_{k,slow}+N_{k,fast}=N_0\frac{v_{k,slow}}{c_{slow}}\mathbf{\hat{k}\cdot\hat{\xi}_{slow}}+N_0\frac{v_{k,fast}}{c_{fast}}\mathbf{\hat{k}\cdot\hat{\xi}_{fast}}.
\end{equation}

The displacement vectors $\mathbf{\xi}$ are given by

\begin{equation}\label{4} 
\mathbf{\xi_{Alfven}}=\hat{\mathbf{k}}_{\perp}\times\hat{\mathbf{k}}_{\parallel},
\end{equation}

\begin{equation}\label{5} 
\mathbf{\xi_{slow}}\varpropto(-1+\alpha-\sqrt{D})k_{\parallel}\hat{\mathbf{k}}_{\parallel}+(1+\alpha-\sqrt{D}))k_{\perp}\hat{\mathbf{k}}_{\perp},
\end{equation}

\begin{equation}\label{6} 
\mathbf{\xi_{fast}}\varpropto(-1+\alpha+\sqrt{D})k_{\parallel}\hat{\mathbf{k}}_{\parallel}+(1+\alpha+\sqrt{D}))k_{\perp}\hat{\mathbf{k}}_{\perp},
\end{equation}

The subscript '$k$' represents parameters in wavevector space, and ‘$^\wedge$’ represents corresponding unit vectors. In low-$\beta$ regime, phase speeds of fast and slow modes ($c_{fast}$ and $c_{slow}$) are equal to $V_A$ (Alfvén speed) and $a\cos\theta_{kB_0}$ (a represents sound speed), respectively. The parameter $D=(1+\alpha)^2-4\alpha\cos \theta_{kB_0}$, where $\alpha=\frac{a^2}{V^2_A}$ and $\theta_{kB_0}$ is the wave propagation angle between wave vectors and the background magnetic field.

\begin{figure}[ht!]
\plotone{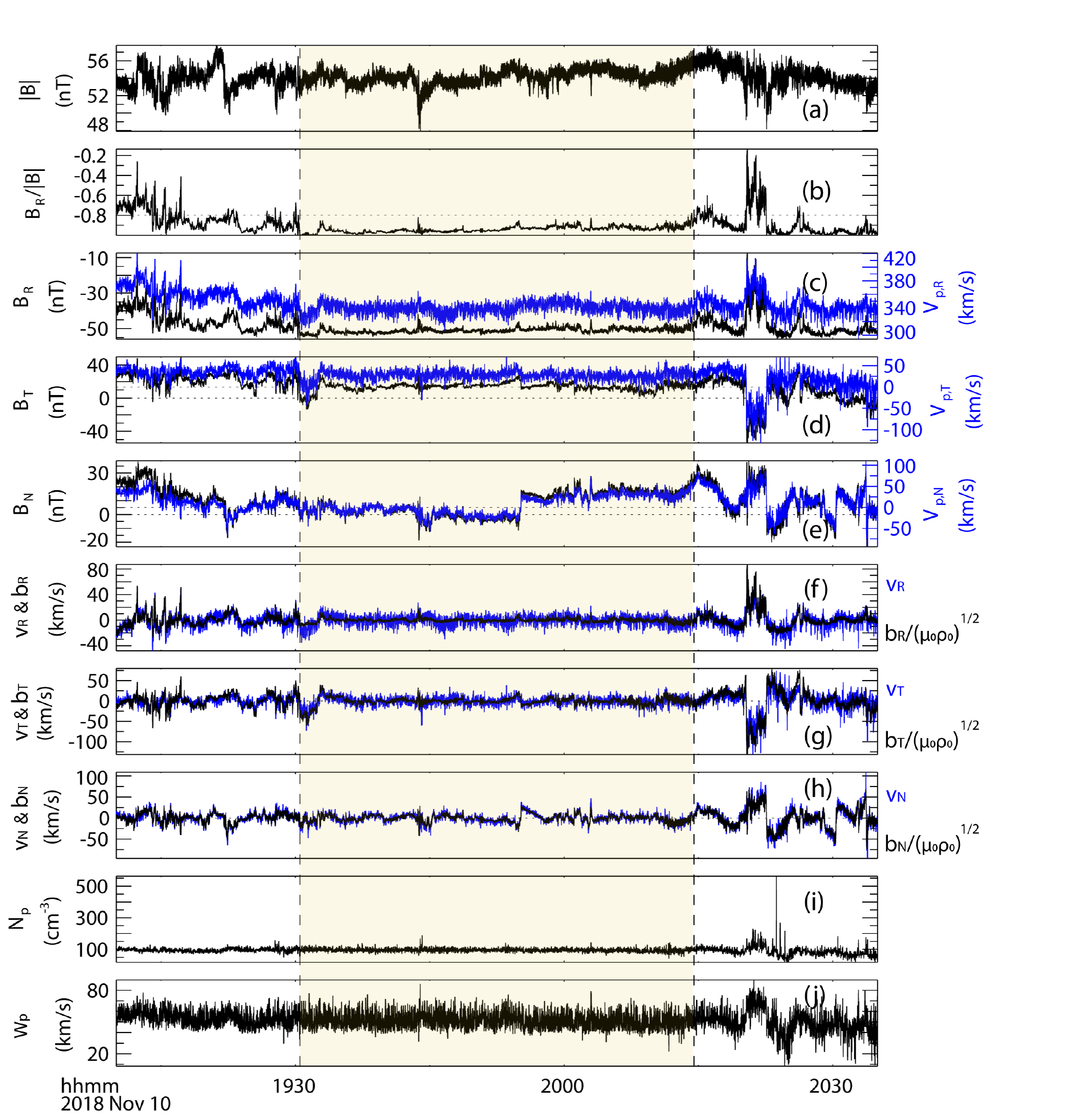}
\caption{PSP observations of a representative case on 10 November 2018 in RTN coordinates within the yellow shaded region. (a) magnetic field magnitude $\mathbf{|B|}$; (b) normalized radial magnetic field $B_R/|\mathbf{B}|>0.8$; (c) radial magnetic field ($B_R$, black) and proton bulk velocity ($V_{p,R}$, blue); (d) the tangential components of magnetic field ($B_T$, black) and proton bulk velocity ($V_{p,T}$, blue); (e) the normal components of magnetic field ($B_N$, black) and proton bulk velocity ($V_{p,N}$, blue); (f-h) high correlations between the magnetic field perturbation $\mathbf{b}$ and proton bulk velocity $\mathbf{v}$, where $\mathbf{b}$ is shown in Alfvén speed units. (i) proton density; (j) proton thermal velocity. RTN: radial tangential normal coordinates.  
\label{fig:Figure1}}
\end{figure}

\begin{figure}[ht!]
\plotone{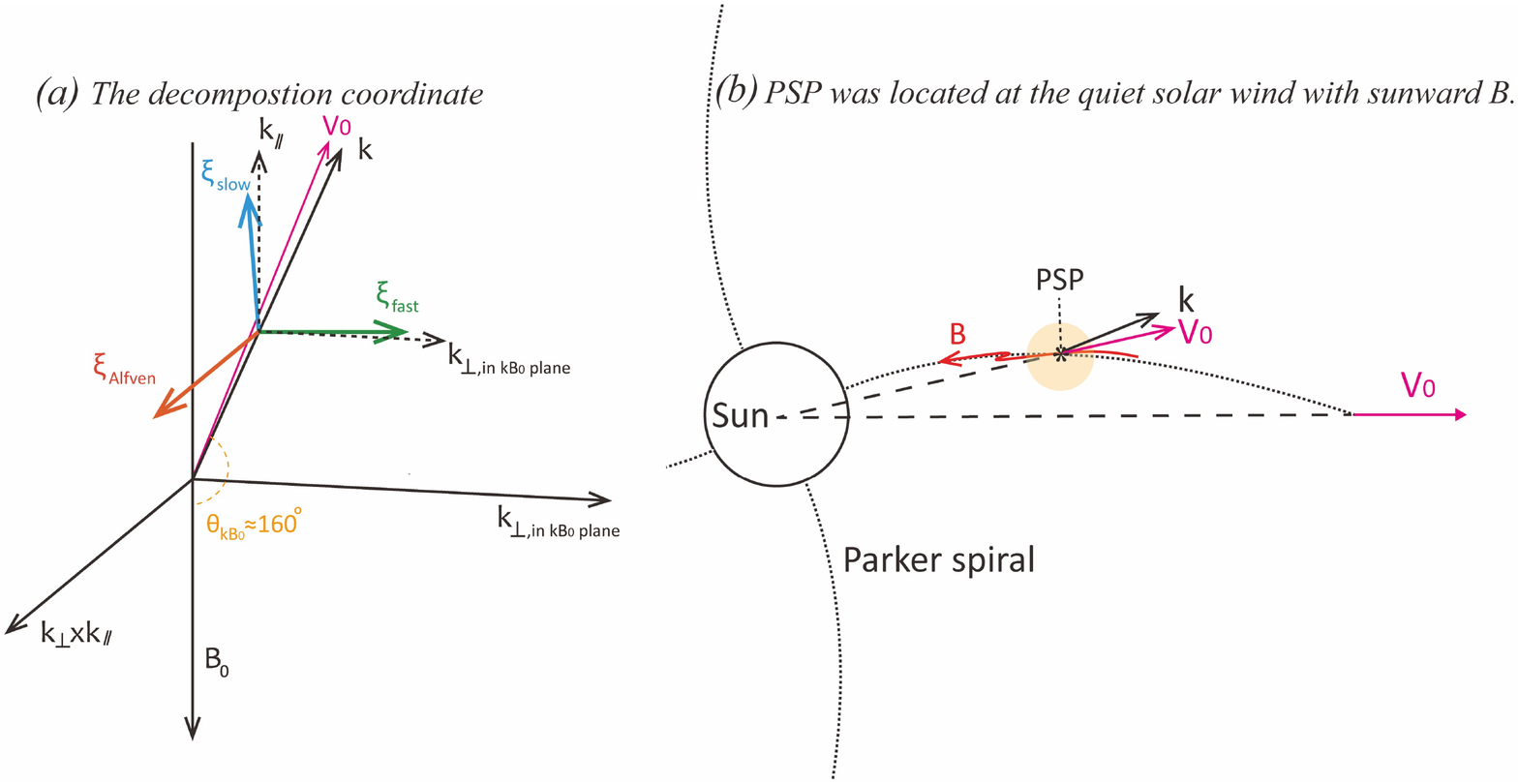}
\caption{Schematic illustrations of (a) the vectors in the decomposition coordinate at $\theta_{kB_0}\sim160^\circ$. The displacement vectors $\mathbf{\xi}$:  $\mathbf{\xi_{Alfven}}$ (red), $\mathbf{\xi_{fast}}$ (green), and  $\mathbf{\xi_{slow}}$ (blue). $\mathbf{\xi_{Alfven}}$ is perpendicular to $kB_0$ plane. $\mathbf{\xi_{fast}}$ and $\mathbf{\xi_{slow}}$ are in $kB_0$ plane. The angle between $\mathbf{V_0}$ and $\mathbf{k}$ is $\sim2^\circ$ in $kB_0$ plane in this event. (b) the relative position between the Sun and PSP;
\label{fig:Figure2}}
\end{figure}

\begin{figure}[ht!]
\plotone{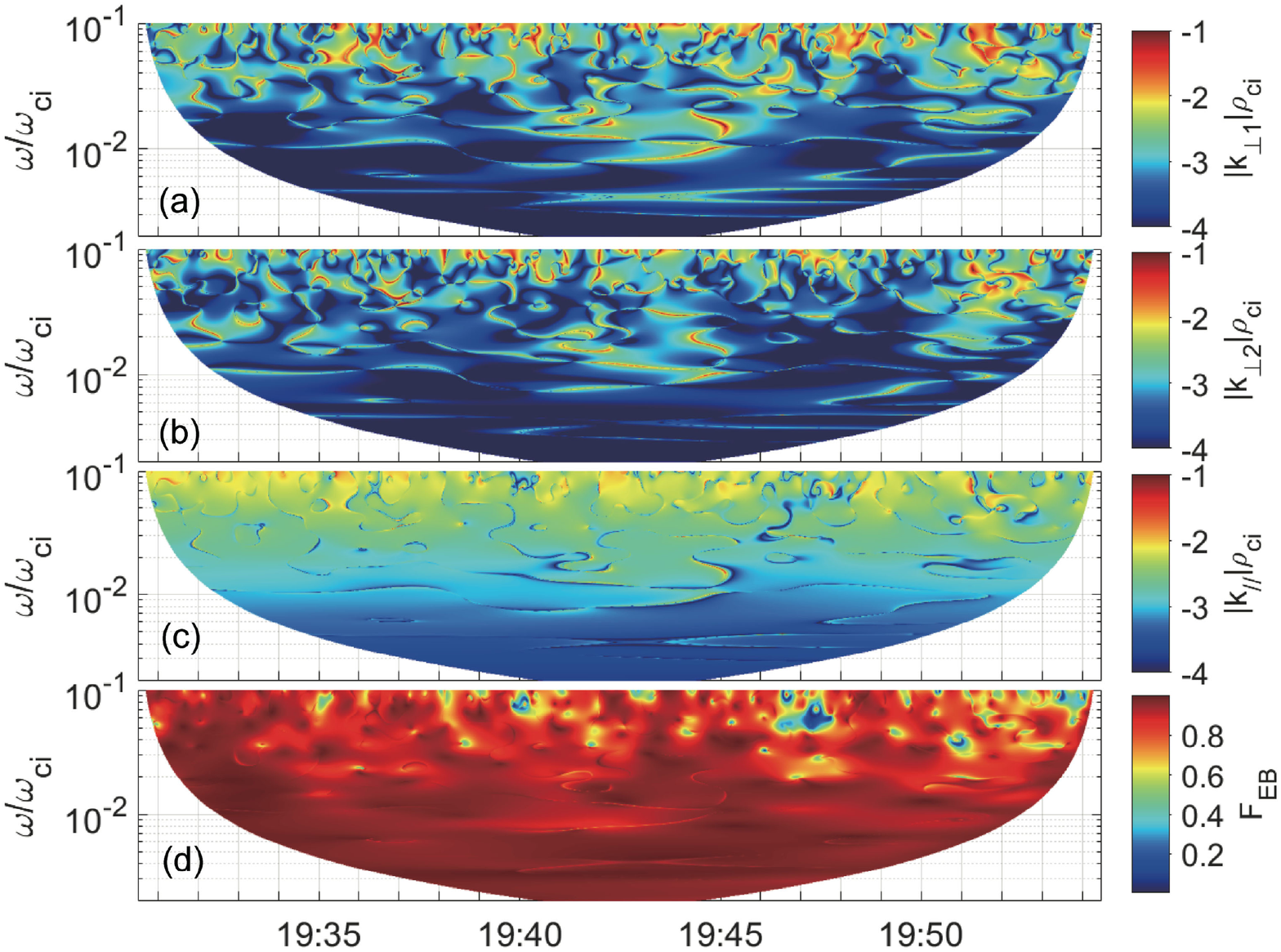}
\caption{(a-c) The spacecraft-frame frequency-time spectrograms of wave vectors $\mathbf{k}$ in the field-aligned coordinates. (d) electromagnetic planarity $F_{EB}$. 
\label{fig:Figure3}}
\end{figure}

\section{Observations and Results}\label{sec:observations}

\subsection{Overview of a representative case} \label{subsec:overview}

Figure \ref{fig:Figure1} shows PSP observations of a representative case of sub-Alfvénic perturbations in low-$\beta$ ($\beta\sim0.18$) plasma at a heliocentric distance of around 0.229 AU on 10 November 2018. The magnetic field and plasma data are shown in RTN coordinates. Figure \ref{fig:Figure1}(a) shows that the magnetic field magnitude $|B|$ is almost constant, indicating weak magnetic compressibility. Figure \ref{fig:Figure1}(b) shows that the normalized radial-field magnetic field $B_R/|\mathbf{B}|$ is less than -0.8 within the yellow shaded region, suggesting that the spacecraft is located at a quiet radial-field solar wind. The corresponding spacecraft relative position is sketched in Figure \ref{fig:Figure2}(b), where the open field lines are emerging from the negative-polarity equatorial coronal hole ($B_R<0$ in Figure \ref{fig:Figure1}(c)) during the first PSP encounter \citep{Bale2019}. In this study, we only focus on the intervals of the sub-Alfvénic non-switchbacks perturbations. In Figure \ref{fig:Figure1}(f-h), the magnetic field perturbation ($\mathbf{b=B-<B>}$) and proton velocity perturbation ($\mathbf{v=V_p-<V_p>}$) present significantly Alfvénic correlations, where $\mathbf{b}$ is shown in Alfvén speed units. During this time interval, $\mathbf{v}\ll\mathbf{V_A}$ (sub-Alfvénic) and $\mathbf{b}\ll\mathbf{B_0}$. Therefore, the perturbations can be considered as a pure superposition of linear modes. Figure \ref{fig:Figure1}(i) shows the proton density with low perturbations. The proton compressibility $C_p$, defined as $C_p=\frac {<n^2>}{N_0^2 }\frac{B^2_0}{<\mathbf{b}\cdot\mathbf{b}>}$ \citep{Gary1986}, is around 0.06, indicating a low level of density compressibility. Figure \ref{fig:Figure1}(j) shows proton thermal velocity with small variations. 

Figure \ref{fig:Figure3}(a-c) illustrates spacecraft-frame frequency-time spectrograms of wave vectors in the field-aligned coordinates determined by the average magnetic field ($\mathbf{B}_0$) and velocity ($\mathbf{V}_0$) during 19:30:30-20:14:30 UT (marked by the yellow shaded region in Figure \ref{fig:Figure1}), where basis vectors $\mathbf{e}_{\perp 1}$, $\mathbf{e}_{\perp 2}$, and $\mathbf{e}_\parallel$ are in $(\mathbf{v}_0\times \mathbf{b}_0)\times \mathbf{b}_0$, $\mathbf{v}_0\times \mathbf{b}_0$, and $\mathbf{b}_0$  directions, respectively ($\mathbf{v}_0$ and $\mathbf{b}_0$ are unit vectors of $\mathbf{V}_0$ and $\mathbf{B}_0$). The wave vectors and frequencies are normalized by average proton gyro-radius and gyro-frequency ($\rho_{ci}$ and $\omega_{ci}$), respectively. Figure \ref{fig:Figure3}(a-c) show spectrograms of absolute values of wave vectors in  $\mathbf{e}_{\perp 1}$, $\mathbf{e}_{\perp 2}$, and $\mathbf{e}_\parallel$ direction, respectively. We can see that $k_\parallel \gg k_\perp$ indicating wave vectors along the magnetic field larger than that across it. Moreover, wave vectors are roughly constant in time in the low-frequency range. In Figure \ref{fig:Figure3}(d), the high electromagnetic planarity ($F_{EB}$ approaching to 1) indicates the presence of a single plane wave, guaranteeing the validity of the SVD method and mode decomposition method \citep{Santolik2003, Cho2003}.

\subsection{The role of each mode in kinetic, magnetic, and density power} \label{subsec:power}
\begin{figure}[ht!]
\plotone{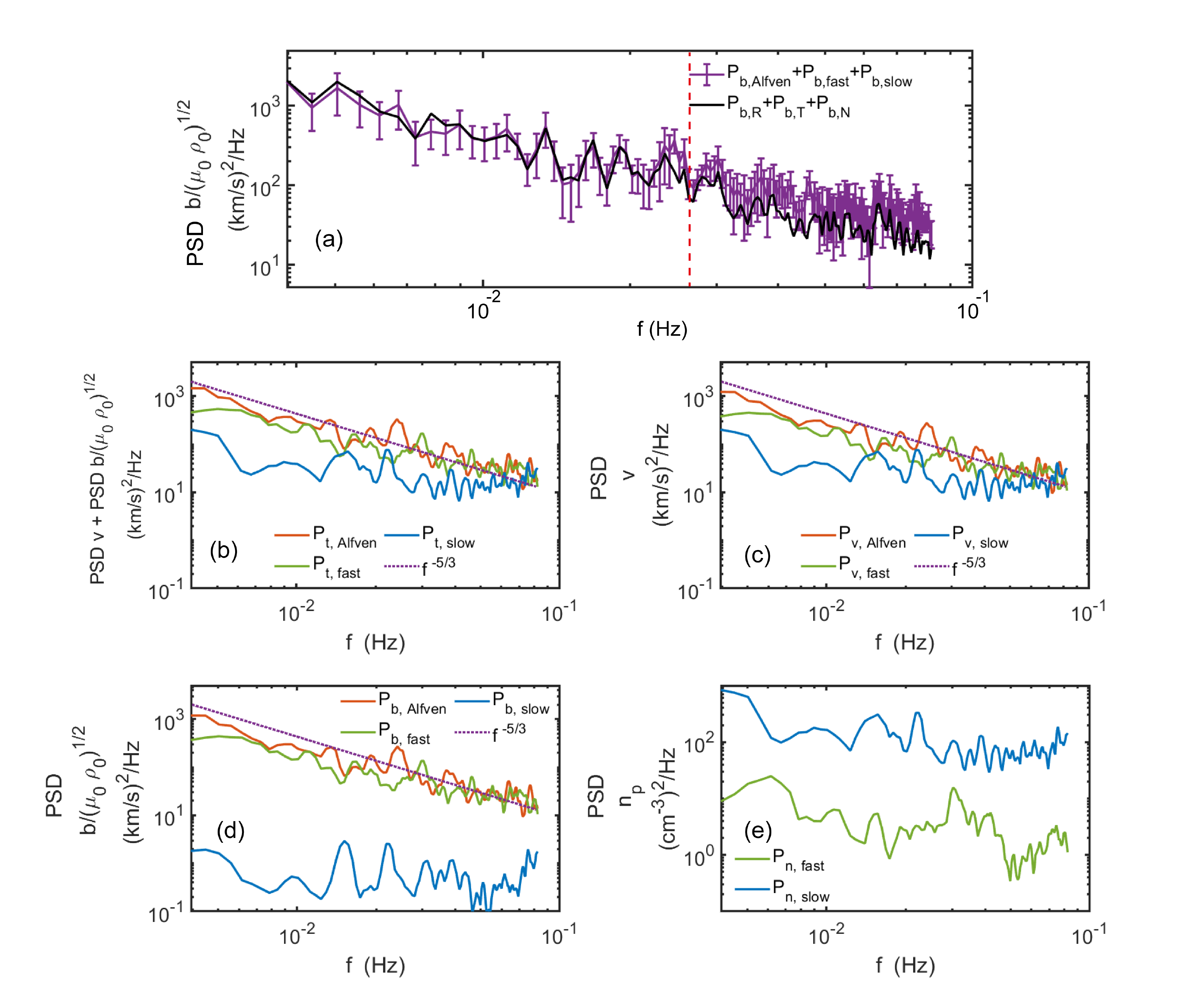}
\caption{Power spectral densities from three MHD modes in the spacecraft frame during 19:30:30-20:14:30 UT. (a) the sum of magnetic power: the observed magnetic power (black; $P_{b,obs}=P_{b,R}+P_{b,T}+P_{b,N}$ and the magnetic power calculated based on the ideal MHD theory (purple; $P_{b,3modes}=P_{b,Alfven}+P_{b,fast}+P_{b,slow}$. The red vertical dashed line represents the frequency $f_{sc}\sim0.026$ Hz. The error bars stand for the standard deviation. (b) total power spectra; (c) kinetic power spectra; (d) magnetic power spectra; (e) proton density power spectra. Red, green, and blue curves represent Alfvén, fast, and slow modes, respectively. The purple dashed lines in (b-d) mark the Kolmogorov-like power law ($f^{-\frac{5}{3}}_{sc}$) as a reference.  
\label{fig:Figure4}}
\end{figure}

To facilitate comparison with direct observations, Figure \ref{fig:Figure4} shows the decomposition results in the spacecraft frame during 19:30:30-20:14:30 UT. To reduce the error, we first set the domain for averaging as a 20-minutes-wide moving window, with a step size of 10 s. Since frequency-time spectrograms of wave vectors and background magnetic fields are roughly constant in time within each window (Figure \ref{fig:Figure1} and \ref{fig:Figure3}), we use temporally averaged wave vectors and magnetic fields to build a new coordinate for mode decomposition. Then, we average the decomposition results overall 20-minutes-wide windows.

According to the definition of Elsasser spectra \citep{Bruno2013} , we calculate magnetic energy $E_b=\frac{1}{2}\mathbf{b}^2$, kinetic energy $E_v=\frac{1}{2}\mathbf{v}^2$, and total energy $E_t=E_b+E_v$. As described in Section \ref{subsec:decompose}, the time series of velocity perturbation is transformed by FFT with three-point smoothing. Combined with the SVD method, the velocity perturbation ($\mathbf{v}_k$) as a function of wave vectors is projected onto the corresponding displacement vectors $\mathbf{\xi}$ ($\mathbf{\xi_{Alfven}}$, $\mathbf{\xi_{fast}}$, and $\mathbf{\xi_{slow}}$, see Figure \ref{fig:Figure2}(a)). Then the magnetic field perturbation and density are calculated based on the ideal MHD theory.

Figure \ref{fig:Figure4}(a) shows the comparison between the sum of the magnetic power of three MHD modes calculated based on the ideal MHD theory (purple;$P_{b,3modes}=P_{b,Alfven}+P_{b,fast}+P_{b,slow}$) and the magnetic power directly obtained by FFT on magnetic field data (black; $P_{b,obs}=P_{b,R}+P_{b,T}+P_{b,N}$) in the spacecraft frame, where the error bars denote the standard deviation. $P_{b,3modes}$ and $P_{b,obs}$ are in good agreement when the frequency $f_{sc}$ is less than $\sim$0.026 Hz (marked by the red vertical dashed line in Figure \ref{fig:Figure4}(a)), confirming the validity of our mode decomposition procedure. The MHD frequency range can be considered as $f_{sc}\sim[0.004,0.026]$ Hz.

The energy fraction of different MHD modes in the inertial range is estimated by
\begin{equation}\label{7} 
P_{MEm}=100\%\times\frac{\sum_f|{U_{k,m}(\mathbf{b})}|^2}
{\sum_{k,m}(|U_{k,m}(\mathbf{b})|^2+|U_{k,m}(\mathbf{v})|^2)}
\end{equation}

\begin{equation}\label{8} 
P_{KEm}=100\%\times\frac{\sum_f|{U_{k,m}(\mathbf{v})}|^2}
{\sum_{k,m}(|U_{k,m}(\mathbf{b})|^2+|U_{k,m}(\mathbf{v})|^2)}
\end{equation}

Here, $P_{MEm}$ and $P_{KEm}$ represent magnetic and kinetic energy fractions, respectively. The spectra $U_{k,m}(\mathbf{v})$  are the energy density for each mode, calculated by velocity perturbation ($\mathbf{v}$) along the corresponding displacement vectors at each wavevector scale. The spectra $U_{k,m}(\mathbf{b})$ are calculated based on Equation (2), where $\mathbf{b}$ is normalized by $\sqrt{\mu_0\rho_0}$ \citep{Bruno2013}. For Alfvén and fast modes, the kinetic energy fraction is roughly equal to the magnetic fraction, where $P_{MEA}\sim P_{KEA}\sim31\%$, $P_{KEf}\sim17\%$, and $P_{MEf}\sim16\%$. For slow modes, the kinetic energy fraction $P_{KEs}$ accounts for $\sim5\%$, whereas their magnetic energy fraction $P_{MEs}$ is negligible. 

Figure \ref{fig:Figure4}(b-e) show power spectral densities (PSDs) of each MHD mode in the spacecraft frame, where red, green, and blue curves represent Alfvén, fast, and slow modes, respectively. In Figure \ref{fig:Figure4}(b), total power is dominated by Alfvén modes, in agreement with the significantly Alfvénic features shown in Figure \ref{fig:Figure1}(f-h). Fast modes make a considerable contribution to total power, even comparable to Alfvén modes as the frequency increases. Slow modes play a limited part in total power. In Figure \ref{fig:Figure4}(c), similar proportions are shown in kinetic power. The slow-mode kinetic power is smaller than those of Alfvén- and fast-mode modes but still holds a certain proportion (Figure \ref{fig:Figure4}(c)).

\begin{figure}[ht!]
\plotone{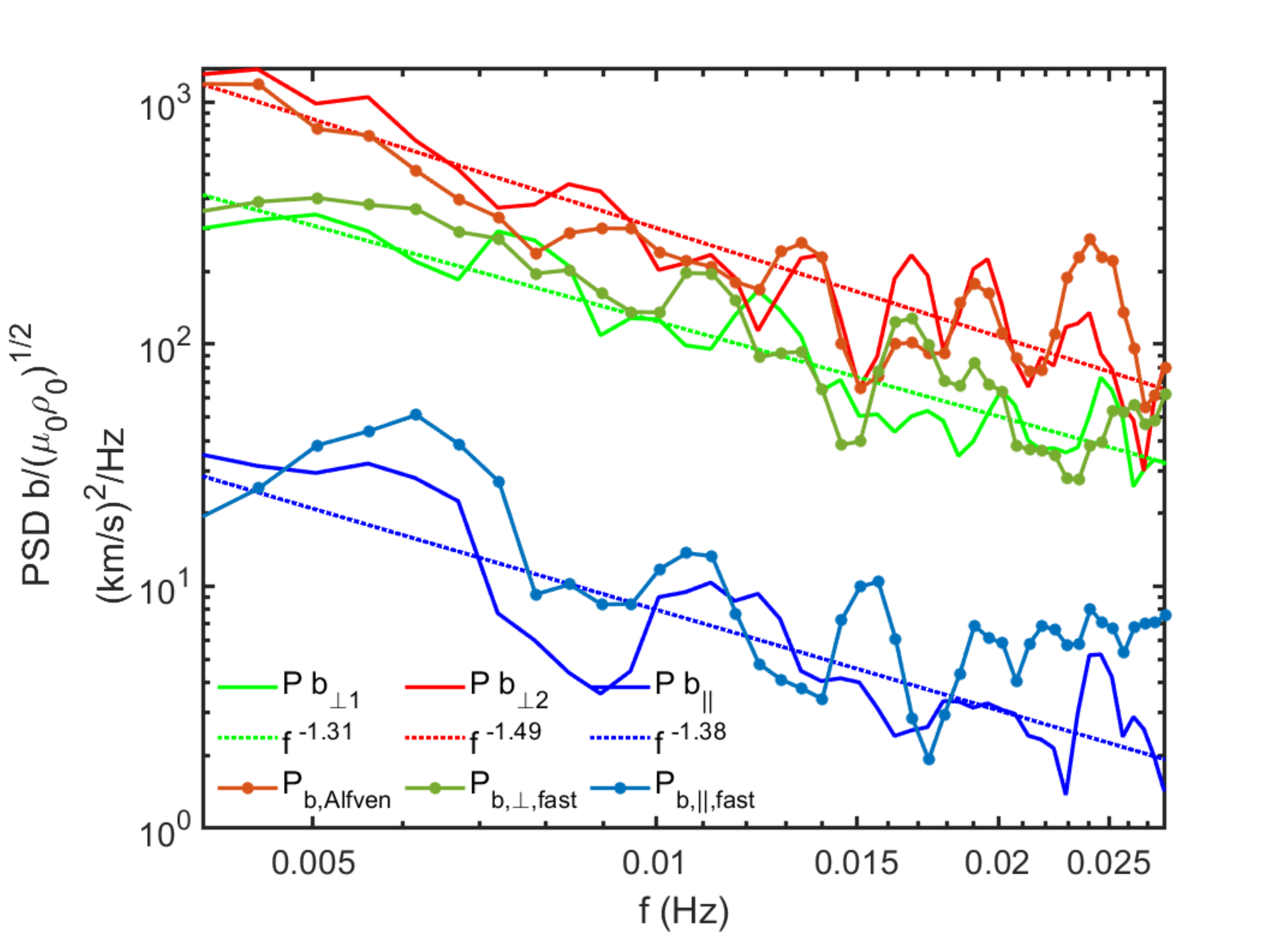}
\caption{The comparisons of magnetic power spectra from direct observations and mode decomposition result from 19:30:30-20:14:30 UT.
\label{fig:Figure5}}
\end{figure}

Figure \ref{fig:Figure4}(d) shows that magnetic power is almost provided by Alfvén and fast modes, whereas slow-mode contributions can be negligible. Figure \ref{fig:Figure5} compares magnetic power spectra from direct observations with mode decomposition results (from Alfvén and fast modes) at the MHD scale. Magnetic field data are transformed into field-aligned coordinates (the same as Figure \ref{fig:Figure3}). The PSDs of  ${b}_{\perp1}$ and ${b}_{\perp2}$ components of the magnetic field perturbation are much larger than those of  ${b}_{\parallel}$ component, indicating that magnetic field perturbations are mostly contained perpendicular to $\mathbf{B}_0$. Moreover, $P_{b\perp2}$ with the scaling $\sim-$1.49 is comparable to Alfvén-mode magnetic power. $P_{b\perp1}$ and  $P_{b\parallel}$ are comparable to the perpendicular and parallel components of fast-mode magnetic power, respectively. Since the angle between $\mathbf{V_0}$ and $\mathbf{k}$ is $\sim2^\circ$, and $\mathbf{V_0}$ is almost in $kB_0$ plane in this event (sketched in Figure \ref{fig:Figure2}(a)), the $V_0B_0$ plane and $kB_0$ plane are roughly coplanar. The Alfvén-mode magnetic field perturbations are perpendicular to $kB_0$ plane, leading to their magnetic power mainly along $\mathbf{e}_{\perp2}$ direction. Similarly, fast-mode magnetic field perturbations are in $kB_0$ plane, which explains the consistency between $P_{b\perp1}$ and $P_{b,\perp,fast}$.

Since slow-mode magnetic power is negligible and slow-mode density power is much larger than that of fast modes (Figure \ref{fig:Figure4}(e)), we speculate that fast modes dominate magnetic compressibility, and slow modes dominate density compressibility.

\begin{figure}[ht!]
\plotone{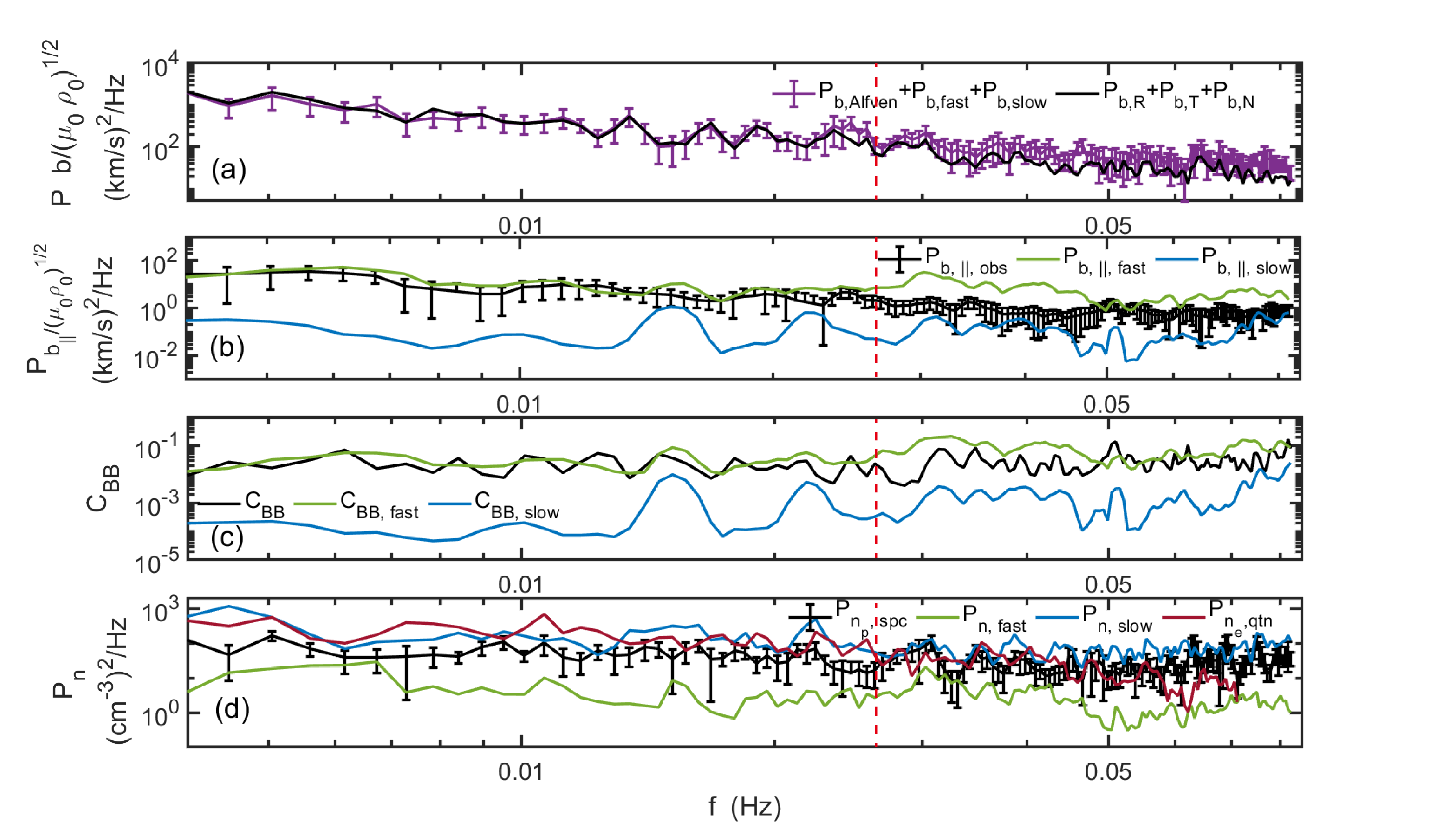}
\caption{Comparison of compressible components of different modes during 19:30:30-20:14:30 UT in the spacecraft frame. (a) the sum of magnetic power in Alfvén units: the observed magnetic power (black; $P_{b,obs}=P_{b,R}+P_{b,T}+P_{b,N}$) and the magnetic power calculated based on the ideal MHD theory (purple; $P_{b,3modes}=P_{b,Alfven}+P_{b,fast}+P_{b,slow}$). (b) spectra of parallel component of magnetic power in Alfvén speed unit. (c) spectra of magnetic compressibility $C_{BB}=(\delta|\mathbf{B}|/|\delta\mathbf{B}|)^2$, $C_{BB,fast}=\frac{P_{b,\parallel,fast}}{P_{b,3modes}}$, and $C_{BB,slow}=\frac{P_{b,\parallel,slow}}{P_{b,3modes}}$. (d) density power spectra: proton density power $P_{n_p,spc}$ from SPC, electron density power $P_{n_e,qtn}$ from QTN, and density power provided by fast modes $P_{n,fast}$ and slow modes $P_{n,slow}$. The error bars represent the standard deviation. The red vertical dashed line represents $f_{sc}\sim$0.026 Hz.
\label{fig:Figure6}}
\end{figure}

We compare PSDs of the compressible component from the direct observations with the decomposition results to verify our speculation on magnetic and density compressibility. In Figure \ref{fig:Figure6}(b), the parallel component of magnetic power ($P_{b,\parallel,obs}$) is obtained by FFT on magnetic field data from MAG, and fast-mode parallel magnetic power ($P_{b,\parallel,fast}$) is calculated based on the ideal MHD theory. $P_{b,\parallel,obs}$ is comparable to $P_{b,\parallel,fast}$ in the MHD regime, indicating that the parallel component of the magnetic field perturbation ($\delta B_\parallel$) is mainly provided by fast modes. The earlier assumption of $\delta{|\mathbf{B}|}\approx\delta B_\parallel$ from slow-modes is invalid in the radial-field solar wind. Moreover, Figure \ref{fig:Figure6}(c) shows the comparison of magnetic compressibility. Fast-mode (slow-mode) magnetic compressibility is defined as the ratio of fast-mode (slow-mode) parallel magnetic power to total magnetic power: $C_{BB,fast}=\frac{P_{b,\parallel,fast}}{P_{b,3modes}}$ ($C_{BB,slow}=\frac{P_{b,\parallel,slow}}{P_{b,3modes}}$). The magnetic compressibility $C_{BB}=(\delta|\mathbf{B}|/|\delta\mathbf{B}|)^2$ is comparable to $C_{BB,fast}$ and much larger than $C_{BB,slow}$, confirming that fast modes dominate magnetic compressibility.

In Figure \ref{fig:Figure6}(d), we compare PSDs of fluctuating density with those provided by fast and slow modes calculated based on the ideal MHD theory (black, proton density power $P_{n_p,spc}$ from SPC; red, electron density power $P_{n_e,qtn}$ from QTN; green, density power provided by fast modes $P_{n,fast}$; blue, density power provided by slow modes $P_{n,slow}$). Due to the relatively low time resolution of electron density data ($\sim$7 s), we obtain the PSD of electron fluctuating density by global FFT with three-point smoothing rather than average moving windows used in proton data. Figure \ref{fig:Figure6}(d) shows that both $P_{n_p,spc}$ and $P_{n_e,qtn}$ are much larger than $P_{n,fast}$, indicating that fast modes cannot provide enough density perturbations. Therefore, it is likely that slow modes provide density perturbations from qualitative aspects. Moreover, $P_{n,slow}$ is systematically larger than $P_{n_p,spc}$ whereas at the same order of magnitude as $P_{n_e,qtn}$, further quantitatively verifying that density perturbations are mainly provided by slow modes. Given the consistency in magnetic power and electron density power, the systematic low $P_{n_p,spc}$ is likely caused by SPC data measurement errors \citep{Liu2021}. Therefore, the mode decomposition method used in this study provides an auxiliary for the calibration of density data.

\subsection{Dispersion relations and collisionless damping of compressible modes} \label{subsec:damping}

\begin{figure}[ht!]
\plotone{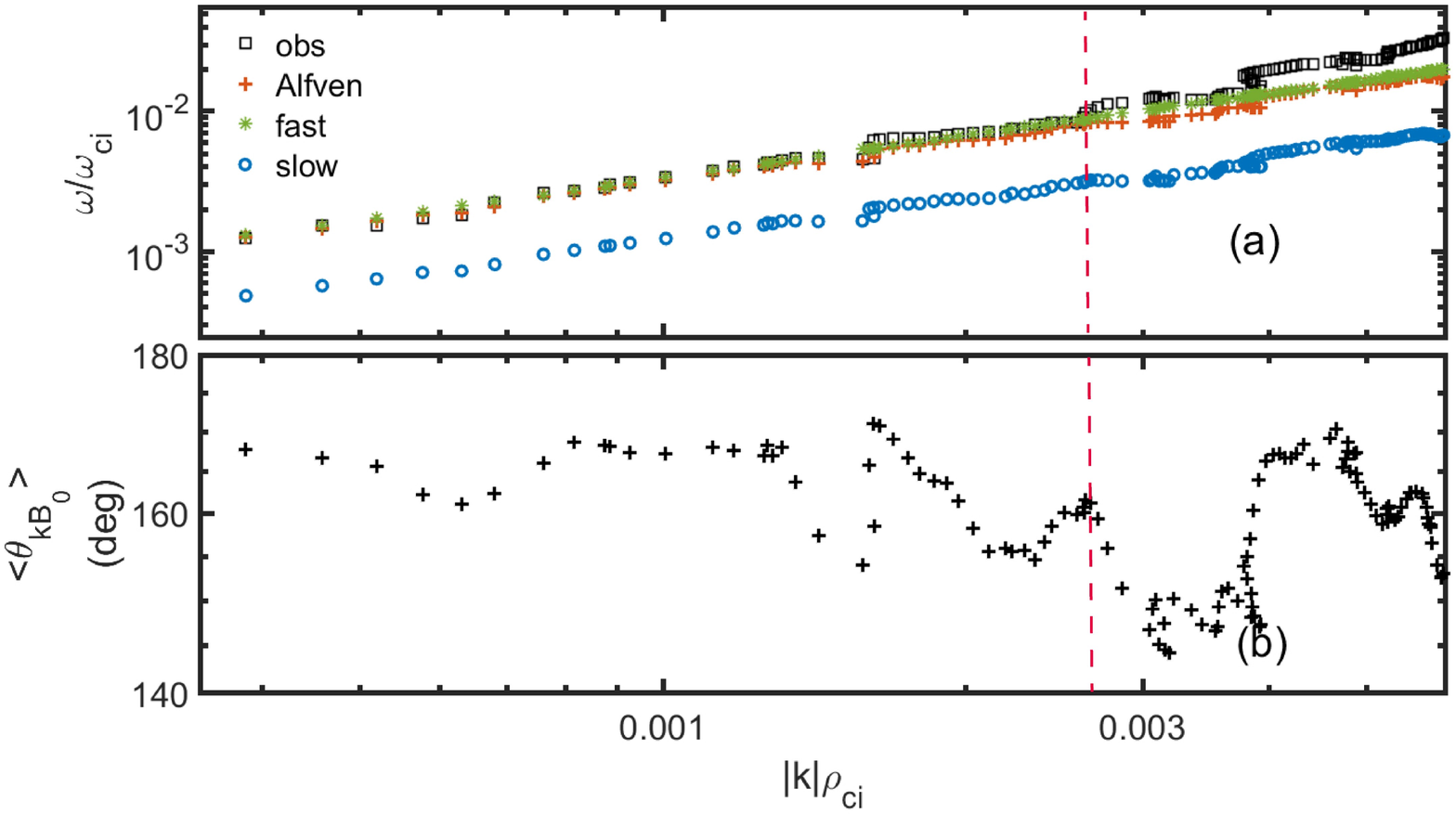}
\caption{(a) Dispersion relations in the plasma flow frame during 19:30:30-20:14:30 UT. (b) The interval-averaged wave propagation angle. The wave vectors and frequencies are normalized by average proton gyro-radius and gyro-frequency ($\rho_{ci}$ and $\omega_{ci}$), respectively. The red vertical dashed line corresponds to $\frac{\omega}{\omega_{ci}}\sim0.083$.
\label{fig:Figure7}}
\end{figure}

Since the propagation time of wave packet is much less than nonlinear interaction time ($\frac{l}{V_A}\ll\frac{l}{v}$, where $l$ is the characteristic length scale), and $k_\parallel \gg k_\perp \sim 0$, perturbations exhibit more wave-like characteristics than turbulence. To investigate the propagations of MHD modes, we analyze their dispersion relations. Firstly, according to the Doppler shift, the observed frequency in the plasma flow frame can be obtained by $\omega_{pf}=\omega_{sc}-<\mathbf{k}>\cdot<\mathbf{V_p}>$, where $\omega_{sc}=2\pi f_{sc}$ is the frequency in the spacecraft frame, and $<>$ represents the time average. The observed frequency $\omega_{pf}$ corresponds to the dominant frequency of perturbations at each wavevector scale.  Secondly, based on ideal MHD theory, the wave phase velocity is given by

\begin{equation}\label{9}
V_{ph,Alfven}=V_A\cos \theta_{kB_0}
\end{equation}

\begin{equation}\label{10}
V^2_{ph,\pm}=\frac{1}{2}\{(a^2+V^2_A)\pm [(a^2+V^2_A)^2-4a^2V^2_A\cos ^2\theta_{kB_0}]^{\frac{1}{2}}\}
\end{equation}

Here, ‘+’ and ‘$-$’ stand for fast and slow modes, respectively \citep{Hollweg1975}. The theoretical frequency is calculated by $\omega_m=kV_{ph,m}$, where the subscript $m$ represents Alfvén, fast, and slow modes. 

Figure \ref{fig:Figure7}(a) displays the comparisons between the observed dispersion relation (black) and theoretical ones for Alfvén (red), fast (green), and slow (blue) modes in the plasma flow frame during 19:30:30-20:14:30 UT. The wave vectors and frequencies are normalized by average proton gyro-radius and gyro-frequency ($\rho_{ci}$ and $\omega_{ci}$), respectively. The dispersion relations of Alfvén- and fast-mode are similar because of $k_\parallel \gg k_\perp$ and the low plasma $\beta$ value. Both dispersion relations are in reasonable agreement with the observed results at $\frac{\omega}{\omega_{ci}}<0.083$, indicating that Alfvén and fast modes dominate the MHD perturbations. The existence of the dominant dispersion relation across most of the spectrum further proves the validity of the SVD approach. Only when $\frac{\omega}{\omega_{ci}}>0.083$, theoretical dispersion relations deviate from observations. The plasma-frame frequency $\frac{\omega}{\omega_{ci}}\sim0.083$ corresponds to $f_{sc}\sim$0.026 Hz in the spacecraft frame, consistent with the MHD frequency range identified by comparing magnetic power in Figure \ref{fig:Figure4}(a).

Figure \ref{fig:Figure7}(b) shows that the wave propagation angle is close to $180^\circ$, indicating the waves propagate roughly antiparallel to the background magnetic field. Due to average $<\theta_{kB_0}>\sim163^\circ$ at the MHD scale, Alfvén-mode phase velocity ($V_{ph,Alfven}$) is roughly equal to that of fast mode ($V_{ph,fast}$), explaining the similar dispersion relations between fast and Alfvén modes.

\begin{figure}[ht!]
\plotone{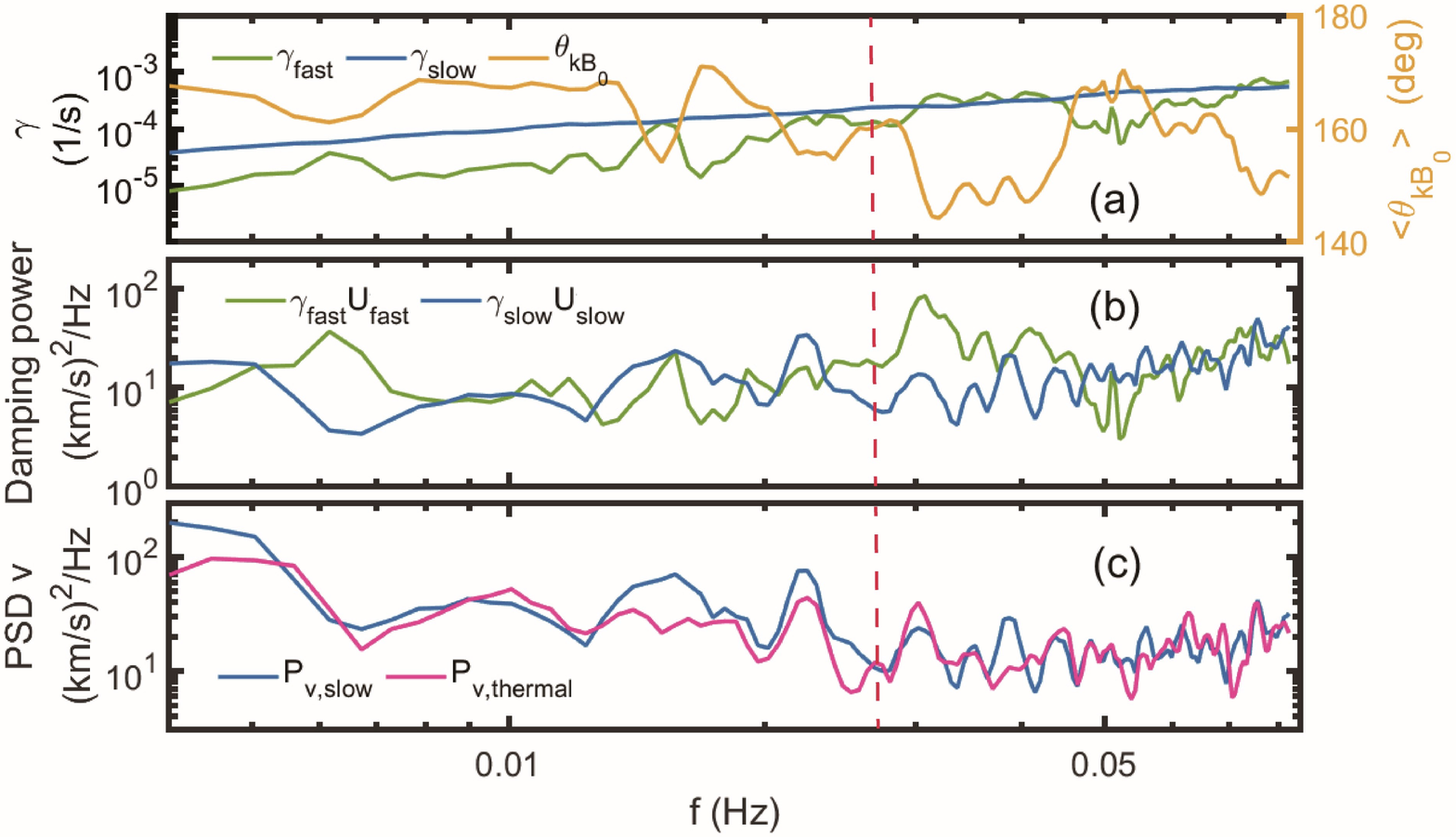}
\caption{(a) Fast-mode damping rate $\gamma_{fast}$, slow-mode damping rate $\gamma_{slow}$, and the interval-averaged wave propagation angle. (b) Wave energy damping power. (c) Slow-mode kinetic power spectral density ($P_{v,slow}$) and the proton thermal fluctuating kinetic PSD (pink; $P_{v,thermal}$). The red vertical dashed line markes $f_{sc}\sim$0.026 Hz.
\label{fig:Figure8}}
\end{figure}

We investigate the possible effects of collisionless damping of each mode on solar wind heating in Figure \ref{fig:Figure8}. In contrast to weakly damped Alfvén modes, compressible modes undergo an intense collisionless wave-damping process, contributing to solar wind heating. The damping rate of fast modes of frequency $\omega$ for $\beta\ll1$ and $\theta_{kB_0}\sim1$ \citep{Ginzburg1961, Yan2004, Petrosian2006} is given by

\begin{equation}\label{11}
\gamma_{fast}=\omega\frac{\sqrt{\pi\beta}}{4}\frac{\sin ^2\theta_{kB_0}}{\cos \theta_{kB_0}}[\sqrt{\frac{m_e}{m_p}}exp(- \frac{m_e}{m_p\beta\cos ^2\theta_{kB_0}})+5exp(- \frac{1}{\beta\cos ^2\theta_{kB_0}})].
\end{equation}

The kinetic damping rate of slow modes \citep{Galeev1983} is given by

\begin{equation}\label{12}
\gamma_{slow}=\frac{|k|a}{2|\cos \theta_{kB_0}|}(\frac{1}{8}\pi\frac{m_e}{m_p})^{\frac{1}{2}}(1-\frac{\cos 2\theta_{kB_0}[(\frac{a^2}{V^2_A})\cos 2\theta_{kB_0}-1]}{[1+\frac{a^4}{V^4_A}-2\frac{a^2}{V^2_A}\cos 2\theta_{kB_0}]^{\frac{1}{2}}}).
\end{equation}
Here, $m_e$ and $m_p$ are the electron and proton mass, respectively.

Figure \ref{fig:Figure8}(a) shows the comparison of the damping rates of compressible modes. For each mode, the damping rate increases with increasing frequency. At the MHD scale, $\gamma_{slow}$ (blue) is systematically larger than $\gamma_{fast}$ (green), suggesting that slow modes suffer more intense collisionless damping. Moreover, fast-mode damping is affected by the wave propagation angle $\theta_{kB_0}$. $\gamma_{fast}$ enlarges when $\theta_{kB_0}$ approaches to $90^\circ$ \citep{Yan2004}, likely because more plasmas can be trapped and interact with fast-mode waves at more significant magnetic compression. In contrast, $\gamma_{slow}$ is scarcely influenced by $\theta_{kB_0}$ under the condition of quasi anti-parallel propagation.

Figure \ref{fig:Figure8}(b) shows wave energy damping power for fast and slow modes, which is defined as the product of the damping rate ($\gamma$) and the mean energy density of the wave ($U$), where $U_{fast}=\frac{1}{2}v^2_{fast}$ and $U_{slow}=\frac{1}{2}v^2_{slow}$. The fast-mode energy damping power $\gamma_{fast}U_{fast}$ accounts for 14.6\%±3.9\% of fast-mode kinetic power ($P_{v,fast}$) in a 95\% confidence interval at the MHD scale. Moreover, the slow-mode energy damping power $\gamma_{slow}U_{slow}$ accounts for 32.2\%±4.0\% of slow-mode kinetic power ($P_{v,slow}$). Since ($P_{v,fast}$) is much larger than ($P_{v,slow}$) (Figure \ref{fig:Figure4}(c)), $\gamma_{fast}U_{fast}$ is roughly comparable to $\gamma_{slow}U_{slow}$. In this process, wave energy can be converted into plasma energy by collisionless damping of compressible modes.

To further quantitively explore their possible contributions to solar wind heating, we calculate the fast-mode energy damping rate by $\epsilon_{fast}=\int\gamma_{fast}U_{fast}df$ and slow-mode energy damping rate by $\epsilon_{slow}=\int\gamma_{slow}U_{fast}df$ at the MHD scale. Results show $\epsilon_{fast}\sim2.66\times10^5 Jkg^{-1}s^{-1}$ and $\epsilon_{slow}\sim2.74\times10^5 Jkg^{-1}s^{-1}$, roughly consistent with the heating rate ($\sim2\times10^{5} Jkg^{-1}s^{-1}$ at the first PSP perihelion) estimated by the global heliospheric simulations \citep{Bandyopadhyay2020}.

Figure \ref{fig:Figure8}(c) shows that slow-mode kinetic PSD (blue; $P_{v,slow}$) is closely related to the proton thermal kinetic PSD (pink; $P_{v,thermal}$), with a correlation coefficient of ~0.80. The proton thermal kinetic energy is defined as $\frac{1}{2}w^2_p$, where $w_p=W_p-<W_p>$ are thermal velocity perturbations of the protons. According to mode decomposition analysis and the continuity equation, slow-mode velocity perturbation (along $\mathbf{\xi}_{slow}$ in Figure \ref{fig:Figure2}(a)) mainly aligns with the background magnetic field and provides most density perturbations. Therefore, we deduce that slow modes may modulate the motion of protons, resulting in thermal energy variations and the inhomogeneous temperature of the plasma.

\section{Discussion} \label{sec:Discussion}
During the first PSP encounter, we identified 15 events (see Table \ref{tab:messier}) based on the selection criteria described in Section \ref{subsec:criteria}. All the events show similar properties to the representative case presented above, such as a high degree of Alfvénicity, stable wave vectors, low magnetic compressibility mainly provided by fast modes, and low density compressibility primarily resulting from slow modes. All our events propagate anti-sunward based on the directions of wave vectors. One possible explanation for the absence of sunward reflected waves is that we can only identify the stronger one when both sunward and anti-sunward waves exist simultaneously. The spacecraft is so close to the Sun that sunward waves do not have enough time to develop sufficiently, and anti-sunward waves dominate (14.6 times more energy in the anti-sunward waves than sunward waves \citep{Chen2021}). The nonlinear interaction is weak, with waves dominating in one direction, and cannot generate strong turbulence.

Comparing with previous studies on MHD mode composition of solar wind turbulence \citep{Chaston2020, Zhu2020}, we set more stringent criteria to guarantee the validity of the single plane-wave assumption and combine the SVD and mode decomposition method \citep{Cho2003} to perform mode decomposition in wave vector space. In the quiet radial-field solar wind, the energy fraction of Alfvén mode is $\sim45\%-83\%$, roughly comparable to the results of \citet{Chaston2020} outside field reversals ($\sim50\%-60\%$). In contrast with them, slow modes ($\sim1\%-19\%$) occupy a much lower proportion than fast mode ($\sim16\%-43\%$) and could even be negligible in some events. The differences come from the sources of perturbations. Considering the limitation of the plane-wave assumption, we analyze the perturbations in the rigorous radial-field solar wind, wherein fast modes show more Alfvénic characteristics, and compressible modes occupy less proportion than those outside field reversals. The new mode decomposition method, apart from providing with energy fraction, makes it possible to quantitively analyze the role of each MHD mode in kinetic, magnetic, and density power spectra. Thereby the contributions of each mode to magnetic and density compressibility are quantified. Moreover, we determine the collisionless damping of each mode, revealing the role of compressible modes in solar wind heating. These results will help us further understand the nature of solar wind turbulence at the MHD scale. 

The MHD perturbations in the radial-field solar wind typically have a small $\theta_{kB_0}$ (approaching to $0^\circ/180^\circ$). Figure \ref{fig:Figure2}(a) shows a schematic illustration of the vectors in the decomposition coordinate at $\theta_{kB_0}\sim160^\circ$, where the angle between $\mathbf{V_0}$ and $\mathbf{k}$ is around $2^\circ$, and $\mathbf{V_0}$ is almost in $kB_0$ plane. $\hat{\mathbf{k}}_{\perp}(\perp\mathbf{B_0})$), $\hat{\mathbf{k}}_{\parallel}(\parallel\mathbf{B_0})$), and $\hat{\mathbf{k}}_{\perp}\times\hat{\mathbf{k}}_{\parallel}$ are unit vectors along the orientations of coordinate axes. As we can see, the fast-mode fluctuating velocities (along $\mathbf{\xi}_{fast}$) are mainly perpendicular to $\mathbf{B}_0$ in $kB_0$ plane, suggesting that their transverse components dominate. Transverse components of fast modes propagating along $\mathbf{B}_0$ show similar characteristics to Alfvén modes, explaining the observed high-degree Alfvénicity. Besides, slow-mode fluctuating velocities (along  $\mathbf{\xi}_{slow}$) almost align with $\mathbf{B}_0$ in $kB_0$ plane, indicating that slow modes are primarily acoustic.

Furthermore, the magnetic field perturbations are determined based on the linearized induction equation ($\omega\mathbf{b}_k=\mathbf{k}\times(\mathbf{B}_0\times\mathbf{v}_k)$). The density perturbations are related to the angles between $\mathbf{k}$ and $\mathbf{v}_k$, based on the continuity equation (Equation (3)). Thus, it is easy to understand that fast modes provide most of the parallel components of the magnetic field perturbations but only tiny density power. By contrast, slow modes dominate density power whereas only provide a tiny amount of magnetic power. Therefore, both in fast-mode nor slow-mode perturbations, the apparent correlation of $n$ and $b$ cannot be observed in the radial-field solar wind.

 Only when $\theta_{kB_0}$ approaching to $90^\circ$, fast modes can produce significant density perturbations based on Equations (3) and (6), and slow modes can produce significant magnetic field perturbations based on Equations (2) and (5). Owing to the enhancements of fast-mode damping rate with $\theta_{kB_0}$ \citep{Yan2004}, it is difficult to identify fast modes based on the positive correlation of $n$ and $b$. By contrast, highly oblique slow waves are less damped and exist in the form of non-propagating pressure-balanced structures \citep{Verscharen2019}. These findings illustrate why few fast waves are observed, whereas slow waves can be frequently detected based on the anti-correlated density-magnetic field strength \citep{Yao2013, SHI2015}.
 
 All radial-field solar wind events in this study show that fast modes provide most of the parallel components of magnetic power, dominating magnetic compressibility. As the Sun rotates, the flow in the inner heliosphere is still radial and not along the jetline, whereas the magnetic field creates a Parker spiral \citep{Parker1957}, as sketched in Figure \ref{fig:Figure2}(b). The angle between the radial direction and the Parker spiral direction increases with the heliocentric distance thereby. Since the angle between $\mathbf{V}_0$ and $\mathbf{k}$ is small in the radial-field solar wind (e.g., $\theta_{kV_0}\sim2^\circ$ in the presented event), it is reasonable to assume that waves propagate in the flow direction, implying that $\theta_{kB_0}$ enlarges with the heliocentric distance. Based on Equations (2) and (6), fast modes become increasingly compressive with the enlargement of $\theta_{kB_0}$. Thus, we deduce that $C_{BB}$ enhancements with the increasing heliocentric distance are attributed to more enhanced compressive fast modes, and the effects of fast modes on compressibility cannot be neglected.
 
 In the radial-field solar wind, wave energy damping power accounts for a considerable proportion in wave kinetic power, such as 14.6\%±3.9\% for fast modes and 32.2\%±4.0\% for slow modes in the presented event. Wave energy can be released into plasma energy by collisionless damping of compressible modes. As $\theta_{kB_0}$ increases, fast modes suffer more intense damping, whereas slow modes with highly oblique propagation are less subject to weak collisionless damping \citep{Yan2004, Verscharen2019}. Therefore, we speculate that fast modes may play a more critical role in plasma heating with increasing heliocentric distance. The radial evolution of each mode's contributions to compressibility and solar wind heating is beyond the scope of this paper. They will be the subjects of our future studies.
 
 We acknowledge the limitations of the SVD method in combination with the modes decomposition study. The SVD provides a linear mapping relationship between frequency and wave vector, whereas $\mathbf{v}_k$ represents the total velocity perturbations at each wavevector scale. Modes decomposition is performed in the space of wave vector, which is retrieved by SVD method and is the only available one currently from single spacecraft measurement. We assume that $\mathbf{v}_k$ transformed by the SVD method includes all perturbations from the three MHD modes at each wave vector. There is no physical reason for the wave vectors of all three modes to be the same. Nonetheless, this assumption hardly affects the results of mode decomposition of turbulence in the radial-field solar wind. We take event \#15 as an example without loss of generality. The decomposition results show that the fraction of Alfvén modes ($\sim62\%$) is slightly higher than that of fast modes ($\sim33\%$), indicating that both Alfvén and fast modes determine the wave vectors. Moreover, fast modes show similar dispersion relations to Alfvén modes (Figure \ref{fig:Figure7}) in low-$\beta$ limit, suggesting the mixture of Alfvén and fast mode affects little SVD results. As for the slow modes, because they only occupy a minor fraction ($\sim5\%$), SVD cannot determine their propagation direction. Nonetheless, in low-$\beta$ limit, the fluctuating velocity along the magnetic field should be mainly provided by slow modes. Even slow modes propagated in a different direction, our mode decomposition method would provide the lower limit of the contribution of slow modes, and the order of the magnitude would not be affected. Due to the inherent temporal and spatial ambiguities of single-spacecraft measurements, the combination method might not be ideal, but one of the best options available until now.

\section{Summary} \label{sec:Summary}

In this study, we report PSP observations of sub-Alfvénic MHD perturbations in low-$\beta$ radial-field solar wind from 31 October to 12 November 2018. We calculate wave vectors using the SVD method \citep{Santolik2003} and separate MHD perturbations into three types of linear eigenmodes (Alfvén, fast, and slow modes) using the mode decomposition method \citep{Cho2003}. Thereby, our research quantitively analyzes the kinetic, magnetic, and density power spectra of each MHD mode and the possible contributions to magnetic and density compressibility. Moreover, we find that collisionless damping of compressible modes may significantly affect solar wind heating based on the correlation of the wave energy damping rate and the heating rate. The specifics of our findings are summarized below.

\begin{enumerate}
\item The radial-field solar wind turbulence has a high degree of Alfvénicity, with the energy fraction of Alfvén modes dominating ($\sim$45\%-83\%) over those of fast modes ($\sim$16\%-43\%) and slow modes ($\sim$1\%-19\%). 
For Alfvén and fast modes, the kinetic energy fraction is roughly equal to the magnetic energy fraction, i.e., $P_{MEA}\sim P_{KEA}\sim31\%$, $P_{KEf}\sim P_{MEf}\sim17\%$ for event \#15. For slow modes, the kinetic energy fraction $P_{KEs}$ accounts for $\sim$5\% (event \#15), whereas their magnetic energy fraction $P_{MEs}$ is negligible.
\item All our events show that fast modes provide most of the parallel components of the magnetic field, dominating magnetic compressibility. Slow modes provide most of the density perturbations, dominating density compressibility.
\item Slow modes modulate the motion of protons, leading to thermal energy variations and the inhomogeneous temperature of the plasma. The energy damping rate of compressible modes is comparable to the solar wind heating rate from the simulations \citep{Bandyopadhyay2020}.
\end{enumerate}

\begin{longrotatetable}
\begin{deluxetable*}{cccccccccccccc}
\tablenum{1}
\tablecaption{List of identified radial-field solar wind turbulence from 31 October to 12 November 2018\label{tab:messier}}
\tablewidth{0pt}
\tablehead{
\colhead{No.} & \colhead{start time} &  \colhead{end time} &
\colhead{scale} & \colhead{$<\beta>$} &
\colhead{$<a>$} &
\colhead{$<V_A>$} &
\colhead{$\theta_{kB_0}$} & \colhead{$P_{KEA}$} & \colhead{$P_{MEA}$} & \colhead{$P_{KEf}$} & \colhead{$P_{MEf}$} &
\colhead{$P_{KEs}$} & \colhead{$P_{MEs}$}  \\
\colhead{} & \colhead{(UT)} & \colhead{(UT)} &
\colhead{(min)} & \colhead{} & \colhead{(km/s)} & \colhead{(km/s)} & \colhead{(deg)} & \colhead{} & \colhead{} & \colhead{} & \colhead{} & \colhead{} & \colhead{}}
\decimalcolnumbers
\startdata
1 & 2018-11-02/16:14:00 & 2018-11-02/17:22:00 & 68 & 0.19 & 41 & 102 & 167 & 29\% & 28\% & 20\% & 20\% & 3\% & $\ll1\%$  \\
2 & 2018-11-03/04:13:00 & 2018-11-03/05:04:00 & 51 & 0.21 & 41 & 98 & 169 & 35\% & 34\% & 13\% & 13\% & 5\% & $\ll1\%$ \\ 
3 & 2018-11-04/19:13:00 & 2018-11-04/19:58:00 & 45 & 0.28 & 50 & 105 & 170 & 27\% & 27\% &21\% & 21\% & 4\% & $\ll1\%$ \\ 
4 & 2018-11-05/00:22:00 & 2018-11-05/01:23:00 & 61 & 0.26 & 50 & 107 & 170 & 29\% & 28\% &20\% & 20\% & 3\% & $\ll1\%$ \\ 
5 & 2018-11-05/15:58:00 & 2018-11-05/16:31:00 & 33 & 0.28 & 50 & 103 & 169 & 27\% & 26\% &22\% & 21\% & 4\% & $\ll1\%$ \\ 
6 & 2018-11-05/20:05:00 & 2018-11-05/21:03:00 & 58 & 0.16 & 47 & 128 & 161 & 23\% & 22\% &19\% & 19\% & 16\% & $\ll1\%$ \\ 
7 & 2018-11-05/21:35:00 & 2018-11-05/22:58:00 & 83 & 0.18 & 45 & 118 & 169 & 28\% & 27\% &21\% & 20\% & 4\% & $\ll1\%$ \\ 
8 & 2018-11-07/04:57:00 & 2018-11-07/05:22:00 & 25 & 0.17 & 43 & 115 & 170 & 30\% & 28\% &20\% & 19\% & 3\% & $\ll1\%$ \\ 
9 & 2018-11-07/09:20:00 & 2018-11-07/10:13:00 & 53 & 0.17 & 70 & 185 & 158 & 29\% & 28\% &17\% & 16\% & 9\% & $\ll1\%$ \\ 
10 & 2018-11-08/07:03:00 & 2018-11-08/07:59:00 & 56 & 0.16 & 52 & 142 & 169 & 36\% & 34\% &14\% & 14\% & 2\% & $\ll1\%$ \\ 
11 & 2018-11-08/16:03:00 & 2018-11-08/16:39:00 & 36 & 0.10 & 48 & 165 & 160 & 42\% & 41\% &8\% & 8\% & 1\% & $\ll1\%$ \\ 
12 & 2018-11-10/00:31:00 & 2018-11-10/01:20:00 & 49 & 0.42 & 81 & 136 & 159 & 26\% & 24\% &21\% & 19\% & 9\% & 1\% \\ 
13 & 2018-11-10/12:00:00 & 2018-11-10/12:38:00 & 38 & 0.18 & 55 & 143 & 165 & 25\% & 25\% &18\% & 18\% & 14\% & $\ll1\%$ \\ 
14 & 2018-11-10/15:40:00 & 2018-11-10/16:35:00 & 55 & 0.12 & 45 & 143 & 157 & 24\% & 24\% &17\% & 16\% & 19\% & $\ll1\%$ \\ 
15 & 2018-11-10/19:30:30 & 2018-11-10/20:14:30 & 44 & 0.18 & 47 & 119 & 163 & 31\% & 31\% &17\% & 16\% & 5\% & $\ll1\%$ \\ 
\enddata
\end{deluxetable*}
\end{longrotatetable}

\begin{acknowledgments}
We would like to thank the members of the FIELD/SWEAP teams and the PSP spacecraft team. We acknowledge the use of data from FIELDS/PSP (\url{http://research.ssl.berkeley.edu/data/psp/data/sci/fields/l2/}) and SWEAP/PSP (\url{http://sweap.cfa.harvard.edu/pub/data/sci/sweap/}). Data analysis was performed using the IRFU-Matlab analysis package available at \url{https://github.com/irfu/irfu-matlab} and the SPADAS analysis software available at \url{ http://themis.ssl.berkeley.edu}.
\end{acknowledgments}

\bibliography{sample631}{}
\bibliographystyle{aasjournal}

\end{document}